\documentstyle[12pt]{article}
\newcommand{\text}{\rm}
\begin{document}

\title{{\bf Symmetry aspects of fermions coupled to torsion and electromagnetic
Fields}}
\author{{\bf J.L. Boldo$^{a}$}\thanks{{\tt E-mail: jboldo@cce.ufes.br}}{\bf \
\thinspace and C.A.G. Sasaki$^{b}$}\thanks{{\tt E-mail: claudio@ime.uerj.br}}%
\vspace{2mm} \\
{\bf $^{a}$}UFES, Universidade Federal do Esp\'{\i }rito Santo\\
CCE, Departamento de F\'{\i}sica\\
Campus Universit\'{a}rio de Goiabeiras 29060-900\\
Vit\'{o}ria, ES, Brazil\vspace{2mm}\\
{\bf $^{b}$}UERJ, Universidade Estadual do Rio de Janeiro,\\
Departamento de Estruturas Matem\'{a}ticas\\
Instituto de Matem\'{a}tica e Estat\'{\i }stica\\
Rua S\~{a}o Francisco Xavier, 524 \\
20550-013, Maracan\~{a}, Rio de Janeiro, Brazil\vspace{2mm}}
\maketitle

\begin{abstract}
We study and explore the symmetry properties of fermions coupled to
dynamical torsion and electromagnetic fields. The stability of the theory
upon radiative corrections as well as the presence of anomalies are
investigated.


\setcounter{page}{0}\thispagestyle{empty}
\end{abstract}

\vfill\newpage\ \makeatother
\renewcommand{\theequation}{\thesection.\arabic{equation}} %
\renewcommand{\baselinestretch}{2}

\section{\ Introduction}

The study of gauge field theories with torsion has been the object of
intensive investigations in recent years due its possible relation with
string theory - the most promising framework for overcoming the problems of
quantum gravity \cite{Green} - which predicts, along with the metric, a
totally antisymmetric rank-3 tensor field usually associated with torsion.
On the other hand, the direct interaction of torsion with fermionic matter
fields has received considerable attention for a long time for the purpose
of formulating general relativity as a gauge theory \cite{Caroll, Hammond,
Shapiro1, Shapiro}. In fact, torsion is the geometric object that relates
the spin of matter with the geometry of the space-time \cite{DeSabbata}.

As it is well known, in Einstein-Cartan gravity, torsion is determined by
the spin distributions and cannot propagate outside matter; however, there
have been discussed recently some theoretical and phenomenological
implications of the hypothesis that torsion is a propagating field \cite
{Caroll, Shapiro}. In particular, the analysis of the consistency of an
effective quantum field theory of spinors interacting with propagating
torsion and electromagnetic fields has been considered in the works of Ref. 
\cite{Shapiro}; the perturbative analysis was performed at one- and two-loop
orders, indicating severe restrictions for the theory.

The aim of our work is to provide a simple algebraic understanding \cite{P-S}
of these results and an extension of them to all orders of perturbations
theory. We shall show the stability of such a model under radiative
corrections and investigate the possible presence of anomalies, by making
use of the symmetry properties of the theory. We shall consider a situation
where matter interacts with a gravitational field whose torsion fluctuations
dominate over the metric excitations. In practice, this means that we adopt
a space-time as a flat background on which the torsion degrees of freedom
propagate and interacts with matter and gauge fields - it might corresponds
to a physical situation where the torsion field is produced by a
cosmological neutrino sea \cite{Letelier}.

Our results indicate that a consistent effective quantum field theory of
torsion - stable under radiative corrections and anomaly free - requires
non-minimal gravitational coupling between the torsion and the (massless)
fermionic sector. Moreover, it contains neither interaction nor coupling
terms between torsion and electromagnetic gauge fields, while keeping the
initial gauge and discrete symmetries.

\section{The Action}

We consider a Riemann-Cartan space-time where the torsion tensor is defined
by $T_{\mu \nu }^{\,\,\,\,\,\,\,\lambda }=2\Gamma _{[\mu \nu
]}^{\,\,\,\,\,\,\,\,\,\,\lambda }=\Gamma _{\mu \nu }^{\,\,\,\,\,\,\,\lambda
}-\Gamma _{\nu \mu }^{\,\,\,\,\,\,\,\lambda }$ and the affine connection $%
\Gamma _{\mu \nu }^{\,\,\,\,\,\,\,\lambda }$ is expressed through metric and
torsion as $\Gamma _{\mu \nu }^{\,\,\,\,\,\,\,\lambda }=\left\{ _{\mu \nu
}^{\,\,\lambda }\right\} +K_{\mu \nu }^{\,\,\,\,\,\,\lambda }$. Here $%
\left\{ _{\mu \nu }^{\,\,\lambda }\right\} $ is the Christoffel symbol while 
$K_{\mu \nu }^{\,\,\,\,\,\,\lambda }=\frac{1}{2}\left( T_{\mu \nu
}^{\,\,\,\,\,\,\lambda }+T_{\,\,\,\mu \nu }^{\lambda }-T_{\nu \,\,\,\mu
}^{\,\,\lambda }\right) $ is the contorsion tensor.

In this space-time, the action for a massless fermion minimally coupled to
torsion and electromagnetic fields has the form: 
\begin{equation}
\Sigma _{\psi }=\frac{i}{2}\int d^{4}x\sqrt{-g}\left( \bar{\psi}\gamma ^{\mu
}\nabla _{\mu }\psi -\nabla _{\mu }\bar{\psi}\gamma ^{\mu }\psi \right) ,
\label{1}
\end{equation}
where the covariant derivatives of the spinor fields are given by: 
\begin{equation}
\nabla _{\mu }\psi =\partial _{\mu }\psi +ieA_{\mu }\psi +\frac{1}{8}B_{\mu
}^{\,\,\,ab}\left[ \gamma _{a},\gamma _{b}\right] \psi ,  \label{2}
\end{equation}
\[
\nabla _{\mu }\bar{\psi}=\partial _{\mu }\bar{\psi}-ieA_{\mu }\bar{\psi}-%
\frac{1}{8}B_{\mu }^{\,\,\,ab}\bar{\psi}\left[ \gamma _{a},\gamma
_{b}\right] . 
\]
Here, Latin indices refer to frame components. $B_{\mu }^{\,\,\,ab}=\omega
_{\mu }^{\,\,\,ab}+K_{\mu }^{\,\,\,ab}$ are the components of the
spin-connection, which is the gauge field of the local Lorentz group. $%
\omega _{\mu }^{\,\,\,ab}$ is the Riemannian part of the spin-connection: 
\begin{equation}
\omega _{\mu }^{\,\,\,ab}=e_{\mu c}\omega ^{\,\,\,cab}=\frac{1}{2}e_{\mu
c}\left( \Omega ^{cab}+\Omega ^{acb}-\Omega ^{bac}\right) ,  \label{3}
\end{equation}
where $\Omega _{abc}=e_{a}^{\mu }e_{b}^{\nu }\left( \partial _{\mu }e_{\nu
c}-\partial _{\nu }e_{\mu c}\right) $ stands for the rotation coefficients
(see \cite{DeSabbata} for details); $e_{\mu }^{a}$ stands for the vierbeins.

Since the main goal of our work is to study the specific effects of a
propagating torsion and its coupling to fermionic matter and electromagnetic
fields, we restrict our analysis to a flat-metric background ($g_{\mu \nu
}=\eta _{\mu \nu }$). Moreover, in our considerations, we shall be dealing
only with the pseudo-vector component of torsion, $S^{\mu }$, since this is
the only mode of $T_{\mu \nu }^{\,\,\,\,\,\,\,\kappa }$ predicted by string
theory and that survives a minimal coupling to fermions. Thus, we set 
\begin{equation}
T_{\mu \nu \kappa }=\varepsilon _{\mu \nu \kappa \lambda }S^{\lambda }.
\label{4}
\end{equation}

From (\ref{1}) and (\ref{2}), one gets the matter action in presence of
torsion and the electromagnetic fields: 
\begin{equation}
\Sigma _{\psi }=\int d^{4}x\left( i\bar{\psi}\gamma ^{\mu }\partial _{\mu
}\psi -eA_{\mu }\bar{\psi}\gamma ^{\mu }\psi +\alpha S_{\mu }\bar{\psi}%
\gamma ^{5}\gamma ^{\mu }\psi \right) ,  \label{5}
\end{equation}
where $\alpha $ is the coupling constant governing the interaction between
fermions and torsion. In the particular case $\alpha =\frac{3}{4}$, the
coupling is minimal in that it comes out from the spin connection with
torsion present in the covariant derivative of the fermion (\ref{2}). Notice
that the coupling constants $e$ and $\alpha $ are dimensionless. Moreover,
it is important to point out that only the totally antisymmetric component
of the torsion minimally couples to the matter field.

In the works of ref. \cite{J-K}, the authors propose and discuss the
coupling between fermions and a background 4-vector to discuss the issue of
CPT and Lorentz violation in gauge theories. Here, our $S_{\mu }$ is a
pseudo-vector, so that our interaction term preserves CPT.

The action is left invariant under the usual U(1) gauge transformations for
the fermion and the electromagnetic fields: 
\begin{eqnarray}
\psi &\rightarrow &e^{ie\Lambda \left( x\right) }\psi ,  \nonumber \\
\bar{\psi} &\rightarrow &e^{-ie\Lambda \left( x\right) }\bar{\psi},
\label{6} \\
A_{\mu } &\rightarrow &A_{\mu }-\partial _{\mu }\Lambda \left( x\right) . 
\nonumber
\end{eqnarray}
Along with such a symmetry, the theory has also an additional local chiral
invariance for which $S_{\mu }$ behaves as the corresponding gauge field:

\begin{eqnarray}
\psi &\rightarrow &e^{i\alpha \theta \left( x\right) \gamma ^{5}}\psi , 
\nonumber \\
\bar{\psi} &\rightarrow &\bar{\psi}\left( x\right) e^{i\alpha \theta \left(
x\right) \gamma ^{5}},  \label{7} \\
S_{\mu } &\rightarrow &S_{\mu }-\partial _{\mu }\theta \left( x\right) . 
\nonumber
\end{eqnarray}
Here, $\Lambda $ and $\theta $ are respectively scalar and pseudo-scalar
gauge functions. We underline that, in contrast to the gauge transformation (%
\ref{6}), (\ref{7}) does not leave invariant a bilinear mass term, $\bar{\psi%
}\psi $, for the fermion.

In the same way as in QED, the action remains invariant under charge
conjugation, ${\cal C}$, symmetry: 
\begin{eqnarray}
\psi &\rightarrow &\psi ^{c}={\cal C}\left( \bar{\psi}\right)
^{t},\,\,\,\,\,\,\,\,{\cal C}=-\gamma ^{2},  \nonumber \\
A^{\mu } &\rightarrow &-A^{\mu },  \label{12} \\
S^{\mu } &\rightarrow &S^{\mu }.  \nonumber
\end{eqnarray}
Our point of view is that $S^{\mu }$ is even under charge conjugation
symmetry since it might {\it only} mediate, along with the metric, the
gravitational interaction between spinor particles.

Furthermore the theory has an additional rigid gauge invariance, namely: 
\begin{eqnarray}
\psi &\rightarrow &e^{i\omega }\psi ,  \label{13} \\
\bar{\psi} &\rightarrow &e^{-i\omega }\bar{\psi},  \nonumber
\end{eqnarray}
where $\omega $ is a constant parameter.

The kinetic action for the gauge sector, compatible with the local and
discrete symmetries (\ref{6}) - (\ref{12}), is as follows:

\begin{equation}
\Sigma _{g}=-\frac{1}{4}\int d^{4}x\,\left( F^{\mu \nu }F_{\mu \nu }+S^{\mu
\nu }S_{\mu \nu }\right) ,  \label{8}
\end{equation}
where $F^{\mu \nu }=\partial ^{\mu }A^{\nu }-\partial ^{\nu }A^{\mu }$ and $%
S^{\mu \nu }=\partial ^{\mu }S^{\nu }-\partial ^{\nu }S^{\mu }$. \ Here, we
do not consider the gauge invariant torsion-photon coupling terms $%
\varepsilon ^{\mu \nu \kappa \lambda }F_{\mu \nu }S_{\kappa \lambda }$ and $%
F^{\mu \nu }S_{\mu \nu }$ since they both break charge conjugation symmetry (%
\ref{12}). In fact, as we shall see later, this action is stable under
radiative corrections; namely, neither coupling nor interaction terms
between torsion and electromagnetic fields are required in the action in
order to fulfill the stability condition.

Since the theory possesses two local symmetries, one has to perform a gauge
fixing for both, i.e.: 
\begin{equation}
\Sigma _{gf}=-\frac{1}{2}\int d^{4}x\,\left[ \frac{1}{\lambda }\left(
\partial A\right) ^{2}+\frac{1}{\beta }\left( \partial S\right) ^{2}\right] ,
\label{9}
\end{equation}
In view of that, let us start from the following action for fermions
interacting with electromagnetic and torsion fields in a flat background: 
\begin{equation}
\Sigma =\Sigma _{g}+\Sigma _{gf}+\Sigma _{\psi }.  \label{10}
\end{equation}

The (continuous) symmetries properties of this model are translated to a set
of the Ward identities, namely: 
\begin{equation}
W_{A}\Sigma =\frac{1}{\lambda }\partial ^{2}\partial
A,\,\,\,\,\,\,\,\,\,\,\,\,\,W_{S}\Sigma =\frac{1}{\beta }\partial
^{2}\partial S\,\,\,\,\,\,{\rm \mbox{and}\,\,\,\,\,\,\,}W_{rig}\Sigma =0
\label{14}
\end{equation}
where 
\begin{equation}
W_{A}=\partial _{\mu }\frac{\delta }{\delta A_{\mu }}-ie\left( \bar{\Psi}%
\frac{\delta }{\delta \bar{\Psi}}-\frac{\overleftarrow{\delta }}{\delta \Psi 
}\Psi \right) ,  \label{15}
\end{equation}
\begin{equation}
W_{S}=\partial _{\mu }\frac{\delta }{\delta S_{\mu }}+i\alpha \left( \bar{%
\Psi}\gamma ^{5}\frac{\delta }{\delta \bar{\Psi}}+\frac{\overleftarrow{%
\delta }}{\delta \Psi }\gamma ^{5}\Psi \right) ,  \label{16}
\end{equation}
\begin{equation}
W_{rig}=\omega \left( \bar{\Psi}\frac{\delta }{\delta \bar{\Psi}}-\frac{%
\overleftarrow{\delta }}{\delta \Psi }\Psi \right) .  \label{17}
\end{equation}
We point out that, in the same way as the gauge fixing terms, bilinear mass
terms for the abelian gauge fields are allowed since they provide only a
linear breaking of the identities (\ref{15}) and (\ref{16}) - whose linear
structure is preserved by radiative corrections. However, the same
conclusion does not hold for the fermionic sector: a fermionic mass term is
not allowed since it provides a non-linear breaking of the Ward identity $%
W_{S}\Sigma =\frac{1}{\beta }\partial ^{2}\partial S.$

\section{Stability}

Once we have established some symmetry properties of the theory, let us
proceed further by studying the stability of the action under radiative
corrections. Let us consider 
\begin{equation}
\Sigma \rightarrow \Sigma +\varepsilon \bar{\Sigma},  \label{18}
\end{equation}
where $\bar{\Sigma}$ is the most general integrated local polynomial of
dimension bounded by four, due to power-counting renormalizability; it has
the same quantum numbers as $\Sigma $. Moreover, it is invariant under local
and rigid gauge transformations, namely: 
\begin{equation}
W_{A\left( S\right) }\bar{\Sigma}=0\,\,\,\,\,\,\,\,{\rm \mbox{and}}%
\,\,\,\,\,\,W_{rig}\bar{\Sigma}=0.  \label{19}
\end{equation}
The most general counterterm, $\bar{\Sigma}$, respecting the discrete ${\cal %
C}$- symmetry along with the conditions (\ref{19}) is given by: 
\begin{equation}
\bar{\Sigma}=-\frac{Z_{A}}{2}F_{\mu \nu }F^{\mu \nu }-\frac{Z_{S}}{2}S_{\mu
\nu }S^{\mu \nu }+2Z_{\psi }\left( \bar{\psi}\gamma _{\mu }\psi A^{\mu }+%
\bar{\psi}\gamma ^{5}\gamma _{\mu }\psi S^{\mu }+\bar{\psi}\gamma _{\mu
}\partial ^{\mu }\psi \right) ,  \label{21}
\end{equation}
where the $Z$'s are arbitrary coefficients. This corresponds to a
redefinition of the fields and coupling constants of the theory, namely: 
\begin{equation}
\Sigma +\varepsilon \bar{\Sigma}=\Sigma \left( \phi _{0};e_{0},\alpha
_{0},\lambda _{0},\beta _{0}\right) ,  \label{21a}
\end{equation}
with 
\begin{eqnarray}
\phi _{0} &=&\left( 1+\varepsilon Z_{\phi }\right) \phi ,\,\,\,e_{0}=\left(
1-\varepsilon Z_{A}\right) e,\,\,\,\alpha _{0}=\left( 1-\varepsilon
Z_{S}\right) \alpha ,  \label{21b} \\
\lambda _{0} &=&\left( 1+2\varepsilon Z_{A}\right) ,\,\,\,\beta _{0}=\left(
1+2\varepsilon Z_{S}\right) ,  \nonumber
\end{eqnarray}
where $\phi $ stands for $A,\,S$ or $\psi $.

Here some remarks are in order. Notice that the bilinear fermionic mass
term, $\bar{\psi}\psi $, cannot show up in the counterterm while local
chiral symmetry is required. In the same way, the term $\varepsilon ^{\mu
\nu \kappa \lambda }A_{\mu }S_{\nu }F_{\kappa \lambda }$ is not allowed by
the requirement $W_{A}\bar{\Sigma}=0$; it is a (electromagnetic) gauge
invariant one only if torsion behaves as a longitudinal gauge field, namely, 
$S_{\mu }=\partial _{\mu }\phi $ \cite{Shapiro1}.

\section{Anomaly}

Now, we shall investigate if the symmetry properties of the model can be
extended to the quantum level for the vertex function 
\begin{equation}
\Gamma =\Sigma +O\left( \hbar \right) .  \label{22}
\end{equation}

Firstly, we notice that the stable action is also invariant under the
parity, ${\cal P}$, transformation: 
\begin{eqnarray}
x^{\mu } &\rightarrow &x_{\mu },  \nonumber \\
A^{\mu } &\rightarrow &A_{\mu },  \label{11} \\
S^{\mu } &\rightarrow &-S_{\mu },  \nonumber \\
\psi &\rightarrow &\gamma ^{0}\psi ,  \nonumber
\end{eqnarray}
where we emphasize, as usual, the pseudo-vector character of $S^{\mu },$
while $A^{\mu }$ behaves as a vector under ${\cal P}$. When investigating
the possible quantum breaking of the continuous symmetries we shall make use
the fact that the discrete symmetries, including parity, must be preserved
at the quantum level.

The Ward operators (\ref{15}) and (\ref{16}) transform in the following way
under parity and charge conjugation symmetries, respectively: 
\begin{eqnarray}
&&W_{A\left( S\right) }\,\,\rightarrow \,\,+\left( -\right) W_{A\left(
S\right) },  \label{23} \\
&&W_{A\left( S\right) }\,\rightarrow \,-\left( +\right) W_{A\left( S\right)
}.  \nonumber
\end{eqnarray}
Then, according to the quantum action principle, the Slavnov-Taylor
identities (\ref{14}) get the following quantum breaking: 
\begin{equation}
W_{A}\Gamma =\frac{1}{\lambda }\partial ^{2}\partial A\,\,+\Delta
_{A}\,\,\,\,{\rm \mbox{and}}\,\,\,\,\,\,W_{S}\Gamma =\frac{1}{\beta }%
\partial ^{2}\partial S+\Delta _{S},  \label{24}
\end{equation}
for $\Delta _{A\left( S\right) }$ a local polynomial of dimension four; in
view of (\ref{23}) it is even (odd) under parity and odd (even) under charge
conjugation symmetries. Having these conditions in mind, we find out that
the symmetry (\ref{6}) is not anomalous: it is straightforward to show,
resolving the corresponding Wess-Zumino consistency conditions \cite{Rouet},
that $\Delta _{A}$ can be written as a $W_{A}$-variation of a local
polynomial, $\hat{\Delta}$, odd under parity and even under charge
conjugation symmetries, namely: 
\begin{equation}
\Delta _{A}=W_{A}\hat{\Delta}.  \label{25}
\end{equation}
Rewriting the vertex function as $\bar{\Gamma}=\Gamma -\hbar \hat{\Delta},$
the Ward identity, $W_{A}\bar{\Gamma}=\frac{1}{\lambda }\partial
^{2}\partial A$, is recovered. On the other hand, $\Delta _{S}$ cannot be
totally written as a $W_{S}$-variation; its functional form reads as: 
\begin{equation}
\Delta _{S}=a\varepsilon ^{\mu \nu \kappa \lambda }S_{\mu \nu }S_{\kappa
\lambda }+W_{S}\left( ...\right) ,  \label{26}
\end{equation}
then we have a breakdown of the local chiral symmetry by quantum effects -
the so called ABBJ anomaly \cite{Adler} present on spaces with torsion \cite
{Mielke}. However, it is well known that its coefficient, $a$, vanishes
whenever one consider a model involving, for instance, two spinor fields
interacting with torsion, with opposite coupling constants. Thus, in this
case, the gauge transformation (\ref{7}) for the spinor field should be
replaced by $\psi _{i}\rightarrow e^{i\alpha _{i}\theta \left( x\right)
\gamma ^{5}}\psi _{i}$ ($i=1,2$), with $\alpha _{1}+\alpha _{2}=0$ - which
means that one of the $\alpha _{i}$ could not be equal to $\frac{3}{4}$ - in
agreement with the conjecture that a renormalizable model of fermions
coupled to torsion requires non-minimal couplings \cite{Shapiro2}.

\vspace{5mm}

{\Large Acknowledgments}

Thanks are due to J.A. Helay\"{e}l-Neto, O. Piguet and S.P. Sorella for
helpful comments and clarifying discussions. The Conselho Nacional de
Desenvolvimento Cient\'{\i }fico e Tecnol\'{o}gico, CNPq, is acknowledged
for its financial support.

\end{document}